# Containers Orchestration with Cost-Efficient Autoscaling in Cloud Computing Environments

Maria A. Rodriguez and Rajkumar Buyya

Cloud Computing and Distributed Systems (CLOUDS) Laboratory
School of Computing and Information Systems
The University of Melbourne, Australia

**Abstract**

Containers are standalone, self-contained units that package software and its dependencies together. They offer lightweight performance isolation, fast and flexible deployment, and fine-grained resource sharing. They have gained popularity in better application management and deployment in recent years and are being widely used by organizations to deploy their increasingly diverse workloads such as web services, big data, and IoT in either proprietary clusters or cloud data centres. This has led to the emergence of container orchestration platforms, which are designed to manage the deployment of containerized applications in large-scale clusters. The majority of these platforms are tailored to optimize the scheduling of containers on a fixed-sized private cluster but are not enabled to autoscale the size of the cluster nor to consider features specific to public cloud environments. In this work, we propose a comprehensive container resource management approach that has three different objectives. The first one is to optimize the initial placement of containers by efficiently scheduling them on existing resources. The second one is to autoscale the number of resources at runtime based on the current cluster's workload. The third one is a rescheduling mechanism to further support the efficient use of resources by consolidating applications into fewer VMs when possible. Our algorithms are implemented as a plugin-scheduler for Kubernetes platform. We evaluated our framework and the effectiveness of the proposed algorithms on an Australian national cloud infrastructure. Our experiments demonstrate that considerable cost savings can be achieved by dynamically managing the cluster size and placement of applications. We find that our proposed approaches are capable of reducing the cost by 58% when compared to the default Kubernetes scheduler.

1. Introduction

Cloud native architectures are becoming a popular approach to structuring and deploying large-scale distributed applications. Contrary to traditional monolithic architectures, these are composed of several smaller, specialized processes, often referred to as microservices, that interact with each other to provide services to users. Container technologies such as Docker [1] and Linux Containers (LXC) [2] provide a lightweight environment for the deployment of these microservices either in private data centers or in virtual clusters in public cloud environments.

Containers are standalone, self-contained units that package software and its dependencies together. Similar to Virtual Machines (VMs), containers are a virtualization technique that enable the resources of a single compute node to be shared between multiple users and applications. However, while VMs virtualize resources at the hardware-level, containers do so at the operating system-level. They are isolated user-space processes that despite running on a shared operating system (OS), create the illusion of being deployed on their own isolated OS. This makes them a lightweight virtualization approach that enables application environment isolation, fast and flexible deployment, and fine-grained resource sharing. Verma et al. [3] demonstrated for instance how using containers at Google led to an improved resource utilization in terms of the number of machines needed to host a given workload on them.

Containerized applications are deployed on a cluster of compute nodes, rather than on a single machine. Organizations are increasingly relying on this technology to deploy diverse workloads derived from modern-day applications such as web services, big data, and IoT. Containers are also found suitable for hosting HPC microservices [20]. This creates the need for container orchestration middleware such as Kubernetes [4], Docker Swarm [6] and Apache Mesos [7]. These systems are responsible for managing and deploying the heterogeneous distributed applications packaged as containers efficiently on a set of hosts.

Hence, a particularly important problem to address in this context is the scheduling or placement of containerized applications on the available hosts. As applications are submitted for deployment, the orchestration system must place them as fast as possible on one of the available resources while considering the application's specific constraints while aiming to maximize the utilization of the compute resources in order to reduce to the operational cost of the organization. This should also be done while considering factors such as the capacity of the available machines, application performance and Quality of Service (QoS) requirements, fault-tolerance, and energy consumption among others. Although the aforementioned frameworks address this issue to an extent, further research is required in order to better optimize the use of resources under different circumstances and for different application requirements.

In cloud environments, containers and VMs can be used together to provide users a great deal of flexibility in deploying, structuring, and managing their applications. In this case, not only should the number of containers scale to meet the requirements of applications, but the number of available compute resources should also adjust to adequately host the required containers. At any given point in time, a cloud container orchestration system should avoid underutilizing VMs as a cost and potentially energy controlling mechanism. It should also be capable of dynamically adding worker VMs to the cluster in order to avoid a degradation in the applications' performance due to resource overutilization. Therefore, autoscaling the number of VMs is essential to successfully meet the performance goals of containerized applications deployed on public clouds and to reduce the operational cost of leasing the required infrastructure. This increases the complexity of the container placement and scheduling problem mentioned above.

Existing container orchestration frameworks provide bin-packing algorithms to schedule containers on a fixed-sized cluster but are not enabled to autoscale the size of the cluster. Instead, this decision is left to the user or to external frameworks at the platform level. An example of such a scenario is using Kubernetes for the placement of containers and Amazon's autoscaling mechanism to manage the cluster. This may not only be impractical but also inefficient as these external entities have limited information regarding the container workload. As a result, we argue that a cloud-centric container orchestration framework capable of making all of the resource management decisions is essential in successfully optimizing the use of resources in virtualized environments.

Another possible optimization to existing systems is related to rescheduling. In particular, rescheduling for either defragmentation or autoscaling when the workload includes long-running tasks. Regardless of how good the initial placement of these tasks is, the performance will degrade over time as the workload changes. This will lead to an inefficient use of resources in which the load is thinly spread across nodes or the amount of resources in different nodes are not sufficient to run other applications. Rescheduling applications that tolerate a component being shut down and restarted will enable the orchestration system to consolidate and rearrange tasks so that more applications can be deployed on the same number of nodes or some nodes can be shutdown to reduce cost or save energy. Similarly, if more nodes are added to the cluster, being able to reschedule some of the existing applications on the new nodes may be beneficial in the long term.

**Key Contribution:** This paper proposes a comprehensive container resource management algorithm that has three different objectives. The first one is to optimize the initial placement of containers so that the number of worker VMs is minimum and the memory and CPU requirements of the containerized applications are met. The second one is to autoscale the number of worker VMs at runtime based on the current cluster's workload. On one hand, scaling out will enable the current resource demand to be met while reducing the time containers wait to be placed and launched. On the other hand, scaling in will enable applications to be relocated so that underutilized VMs can be shutdown to reduce the infrastructure cost. Finally, a rescheduling mechanism will further support the efficient use of resources by consolidating applications into fewer VMs when possible to either avoid unnecessary scaling out operations or to encourage scale in operations.

The rest of the paper is organized as follows. Section 2 presents prominent container management platforms along other related scheduling algorithms. Section 3 discusses application workload models and system requirements of our orchestration framework. Architecture of the proposed system and its realisation by leveraging Kubernetes is discussed in Section 4; and its design and implementation is discussed in Section 5. Autoscaling, rescheduling, and scheduling methods we proposed are discussed in Section 6. Section 7 presents the evaluation of our framework and the effectiveness of the proposed algorithms on an Australian national cloud infrastructure. Section 8 identifies and discusses new research directions. Finally, the last section summarises and concludes the work.

## 2. Related Work

There have been considerable attempts to build reliable and scalable container management platforms for clusters from the research and open-source communities [21]. These systems are responsible for managing the deployment of general-purpose applications, usually packaged in either Docker or Linux cgroups-based containers. In this section, we briefly present related works in the context of software systems and resource management algorithms for orchestration of containers.

*Software Systems:*

Kubernetes [4] is a prominent open-source container management platform, which is derived from Google's in-house Borg [3] software. Kubernetes orchestrates and manages the lifecycle of heterogeneous containerized applications (such as services and batch jobs) in either virtualized or physical clusters. It provides a default scheduler that assigns each container group (or pod) to available cluster resources filtered by user-defined requirements and ranked based on individually defined application affinities. A detailed evaluation of Kubernetes can be found in [5].

Docker Swarm [6] is the native clustering solution for Docker containers. It provides discovery services and schedules containers onto hosts using predefined strategies and filters. By default, it supports two simple strategies: i) a spread strategy that favors running new containers on least loaded hosts, and ii) a bin packing strategy that favors the most loaded hosts that have enough resources to run the containers. Docker Swarm also employs a filtering mechanism to identify the qualified resources; users can define filters regarding host statuses (such as available resources) and health check results and container configurations (such as resource and file affinities and dependencies to other containers).

Apache Mesos [7] enables cluster sharing among various applications. Different from the previous platforms, Mesos can be viewed as a meta-platform that operates above them. It employs a two-stage scheduling approach. In the first stage, it divides the resources of the cluster and respectively provisions resources to each application via resource offers. Once an application accepts a resource offer, it can proceed to scheduling its tasks on the obtained resources using its own scheduling logic. After that, Mesos actually launches the tasks for the application on the corresponding hosts.

Apache Marathon [8] and Apache Aurora [9] are two popular general-purpose frameworks built on top of Mesos. Marathon is designed to orchestrate long-running services. Because of this, it focuses on providing applications with fault-tolerance and high-availability; Marathon will ensure that launched applications will continue to run even in the presence of node failures. Originally developed by Twitter, Aurora is a scheduler that enables long-running services, cron jobs, and ad-hoc jobs to be deployed in a cluster. Aurora specializes in ensuring that services are kept running continuously and as a result, when machine failures occur, jobs are intelligently rescheduled onto healthy machines. It is worthwhile mentioning that in terms of functionality, Marathon and Aurora are very similar products. However, there are a few differences. The main one is that Marathon handles only service-like jobs. Furthermore, setting up and using Marathon is considered to be simpler than doing so with Aurora; Marathon includes for example a user interface through which users can directly schedule tasks.

YARN [10] is a cluster manager designed to orchestrate Hadoop tasks, although it also supports other frameworks such as Giraph, Spark, and Storm. Each application framework running on top of YARN coordinates their own execution flows and optimizations as they see fit. In YARN, there is a per-cluster resource manager (RM) and an application master (AM) per framework. The AM requests resources from the RM and generates a physical plan from the resources it receives. Hence, YARN's two-level scheduling model is similar to Mesos, but instead of it offering resources to applications, applications request the needed resources from YARN.

Although most of the above systems are mature, the scheduling optimization space can still be further explored. This is especially true in the era of cloud computing, as most existing frameworks ignore many of the inherent features of cloud platforms in favour of assuming a static cluster of resources. As a result, elasticity, resource costs, and pricing and service heterogeneities are ignored. Rescheduling for defragmentation and autoscaling is another feature missing from the aforementioned systems.

*Resource Management Algorithms:*

Apart from investigating the orchestration platforms, we discuss relevant works that focused on contained-based resource scheduling algorithms. Xu et al. [22] proposed a resource scheduling approach for the container-based cloud environments to reduce the response time of users' requests and improve providers' resource utilization, which can obtain an optimal mapping for containers to hosts by using stable machine theory. Zhang et al. [23] designed a novel video surveillance cloud platform based on containers, which applies future workload prediction to achieve fine-grained resource provisioning and ensure quality of service. Kaewkasi et al. [24] applied Ant Colony Optimization to allocate containers to hosts and aimed to balance the resource usage. Kehrer et al. [27] presented a two-phase deployment method for Apache Mesos to achieve a more flexible management of containers. Guerrero et al. [26] proposed a genetic algorithm for Kubernetes platform to optimize container allocation and elasticity management. Their algorithm can improve system performance and reduce network overhead. However, all these approaches focus on the initial placement of containers and do not consider the rescheduling and autoscaling.

Yin et al. [25] built a task-scheduling model by exploiting containers and developed a task-scheduling algorithm to reduce the completion time of tasks in fog computing environment. They also proposed a rescheduling approach to optimize task delays. However, the results are only validated in simulated environment rather than dominant container platforms.

Xu et al. [28] proposed an approach based on containers and brownout, which can dynamically activate/deactivate optional containers to achieve the energy-efficient management of cloud resource while ensuring quality of service. This approach applies autoscaling to reduce the active number of hosts while it doesn't consider the initial placement and rescheduling of containers.

Table 1 summaries the comparison of related works focused on resource management algorithms.

Table 1: Comparison of related algorithmic works

| Work | Objective/Focus | Initial Placement | Rescheduling | Autoscaling | Platform |
| --- | --- | --- | --- | --- | --- |
| Xu et al. [22] | Reduce response time | √ | | | Simulation |
| Zhang et al. [23] | Improve resource utilization | √ | | | Docker |
| Kaewkasi et al. [24] | Balance resource usage | √ | | | Docker |
| Yin et al. [25] | Reduce task delays | √ | √ | | Simulation |
| Guerrero et al. [26] | Reduce network overhead | √ | | | Kubernetes |
| Kehrer et al. [27] | Flexible management | √ | | | Mesos |
| Xu et al. [28] | Reduce energy | | | √ | Docker |
| Our work (this paper) | Reduce costs | √ | √ | √ | Kubernetes |

### 3. Workload Models and System Requirements

We consider a container orchestration framework that is deployed in a virtualized public cloud environment by either an organization or a platform as a service provider. Hence, the framework has access to an unlimited number of VMs that can be provisioned and deprovisioned on-demand. For the initial stages of this work, we assume homogeneous VMs, that is, VMs that have the same price and characteristics and are charged per billing period. The characteristics of VMs are defined in terms of their memory and CPU capacities.

To efficiently utilize resources, instead of running separate clusters of homogeneous containers, organizations prefer to run different types of containerized applications on a shared cluster. As a result, we consider a workload that consists of two different types of continuously arriving containerized tasks. The first type of tasks are long-running services that require high availability and must handle latency-sensitive requests. Examples include user-facing web applications or web services. The second type of tasks are batch jobs. These have a limited lifetime and are more tolerable towards performance fluctuations. Examples include scientific computations or map-reduce jobs. We assume that all tasks define the amount of CPU and memory that they require (i.e., resource requests). In this way and to guarantee these requests are fulfilled, tasks will only be deployed on nodes that have at least their required amount of resources available.

Furthermore, long-running services may be defined by users as moveable. Moveable tasks are those that can tolerate being shut down and restarted on a different node for purposes of rescheduling and autoscaling. This may include stateless web applications or replicated web services for instance.

From a high-level perspective, the problem being addressed in this work is divided in three phases:

*Scheduling:* Place an incoming task on a cluster resource that has at least the amount of CPU and memory resources requested by the task available.

*Rescheduling:* If there is an incoming task that is unschedulable (i.e., no node has the required amount of resources available), then the reschedule will attempt to rearrange the placement of moveable running tasks to make room for the unschedulable task.

*Autoscaling:* This problem is divided into two sub-problems: i) scaling out and ii) scaling in. On one hand, if after attempting to reschedule there is still an unschedulable task, the autoscaler should consider scaling out (provisioning a new VM) in order to increase the cluster's capacity to deploy the task. If there are unused VMs or the cluster resources are being utilized inefficiently, the autoscaler should consider shutting down unused VMs or consolidating tasks to increase the utilization of the cluster and deprovision unnecessary VMs.

The description of these three phases and approach noted is high-level; for each phase different strategies that can be implemented with different objectives and goals. Furthermore, the decision to undertake different actions (e.g., reschedule, scale out, scale in) may be made based on different criteria. For instance, a system may choose not to scale out if a new VM was recently provisioned or to wait until a task has been in the pending queue for a certain amount of time before attempting to reschedule. The goal is to develop a container management system that embeds the scheduling, rescheduling, and autoscaling concepts in a general manner so that specific policies can then be plugged into the system.

4. **System Architecture**

Figure 1 depicts the architecture of the system and its realisation by extending the Kubernetes (K8s) platform. The existing K8s components shown in blue, the extended components are shown in green and red. The components show in red are relevant in the context of integrating or using public Clouds such as Amazon. In the rest of this section we first introduce Kubernetes and the key features relevant to this work followed by an overview of each of the proposed extended components.

4.1. **Kubernetes**

Kubernetes is a framework designed to manage containerized workloads on clusters. The basic building block in Kubernetes is a pod. A pod encapsulates one or more tightly coupled containers that are co-located and share the same set of resources. Pods also encapsulate storage resources, a network IP, and a set of options that govern how the pod's container(s) should run. A pod is designed to run a single instance of an application; in this way multiple pods can be used to scale an application horizontally for example. The amount of CPU, memory, and ephemeral storage a container needs can be specified when creating a pod. This information can then be used by the scheduler to make decisions on pod placement. These compute resources can be specified both as a requested amount or as a limit on the amount the container is allowed to consume.

The default Kubernetes scheduler ensures that the total amount of compute resource requests of all pods placed in a node does not exceed the capacity of the node. This even if the actual resource consumption is very low. The reason behind this is to protect applications against a resource shortage on a node when resource usage later increases (e.g., during a daily peak). If a container exceeds its memory limit, it may be terminated and may be later restarted. If it exceeds its memory request, it may be terminated when the node runs out of memory. Regarding the CPU usage, containers may or may not be allowed to exceed their limits for periods of time, but they will not be killed for this. On the other hand, containers and pods that exceed their storage limit will be evicted.

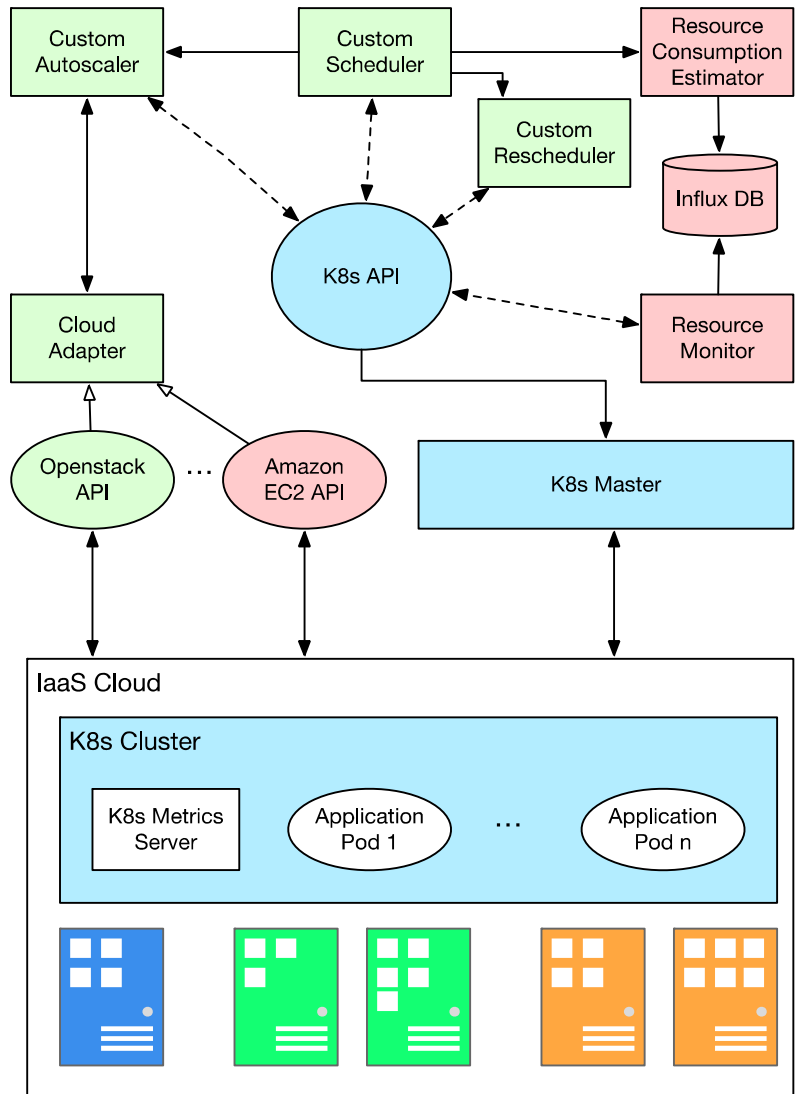

**Figure 1.** Architecture of the prototype system

Other resources (called Extended Resources) can be specified to advertise new node-level resources, their resource accounting is managed by the scheduler to ensure that no more than the available amount is simultaneously allocated to pods.

From a technical perspective, Kubernetes allows for various types of container runtimes to be used, with Docker and rkt natively supported by the platform. More recently, the release of the framework's Container Runtime Interface (CRI) API has enabled Kubernetes to support other container technologies such as containerd [11] and frakti [12], a hypervisor-based container runtime. Furthermore, CRI-O [13], an implementation of the CRI API, currently enables Kubernetes to support any OCI (Open Container Initiative) compliant container runtime such as runc [14]. Also, supporting

Kubernetes in managing the cluster nodes and jobs is etcd [15], an open source, highly-available, distributed key-value store. Specifically, etcd is used to store all of the cluster's data and acts as the single source of truth for all of the framework's components.

Overall, Kubernetes is a highly mature system; it stemmed from ten years of experience at Google with Borg and Omega [16] and is the leading container-based cluster management system with an extensive community-driven support and development base. It provides users with a wide range of options for managing their pods and the way in which they are scheduled, even allowing for pluggable customized schedulers to be easily integrated into the system. To conclude, although Kubernetes' performance and scalability may still not reach the levels of industry-based systems like Borg, as of version 1.10, Kubernetes is capable of supporting clusters of up to 5000 hundred nodes [17], which suits the needs of many organizations nowadays.

### 4.2. Extended Components

A *custom scheduler* interacts with the Kubernetes API to continuously monitor the state of pods in the cluster. A pod corresponds to a single task, either long running or batch. Hence, the scheduler focuses on processing pending pods (i.e., those that need to be scheduled). For each pending pod, a set of suitable resources (i.e., Kubernetes workers) is filtered from the entire cluster pool. One of these resources is then selected and a binding between the pod and the resources is created. This binding leads to Kubernetes running the pod on the chosen node. Once again, it is worthwhile noticing that different policies to select the set of suitable nodes and assign a task to one of them have been implemented and are easily pluggable into the system. These are described in detail in Section 7.

The custom scheduler interacts with the custom rescheduler whenever a pod cannot be placed. The rescheduler then attempts to consolidate moveable pods so that the incoming pod can be placed in an existing node. If the cluster is at capacity and there is no room for the incoming pending pod, then the autoscaler component is invoked in order to determine whether a new VM should be provisioned to place the pod. The autoscaler can then decide the number and type of VMs to launch and achieve this by instructing the Cloud Adapter to create the new instances via the specific IaaS cloud provider API. In the meantime, the unschedulable pod can be left in the scheduling queue in a pending state so that it can be scheduled in a later cycle when the newly provisioned (or recently freed) resources become available. The pod can also be removed from the general scheduling queue so that it can be directly assigned to the newly created node once it is available for use.

The custom scheduler, rescheduler, and autoscaling modules have been developed. The interaction between them is depicted in Algorithm 1. A Cloud Adapter module capable of interacting the Openstack API has also been developed. Other APIs can easily be plugged into the system.

| Algorithm 1 |
|---|
| ```
while the scheduler exit condition is not satisfied
  get all pending tasks
  for each pending task t
    schedule t
      if t cannot be placed
        reschedule
      if rescheduling failed
        scale out
    scale in
``` |

The custom scheduler can also make use of the Resource Consumption Estimator when making decisions. For instance, a batch job may be placed in a node in which not all the requested resources are predicted to be used. Different analytical techniques can be plugged into the Resource

Consumption Estimator to make predictions on the amount of resources consumed by applications. Currently, this functionality has not been integrated into the framework. Instead, we rely on CPU and memory requests defined by tasks to allocate resources. Although this may be beneficial for tasks as they are guaranteed the amount of resources requested even if they are not consumed, it is clearly not optimal in terms of resource usage efficiency. Accurately estimating the amount of resources used for tasks enables the scheduler to make better decisions and use resources opportunistically that would otherwise remain idle.

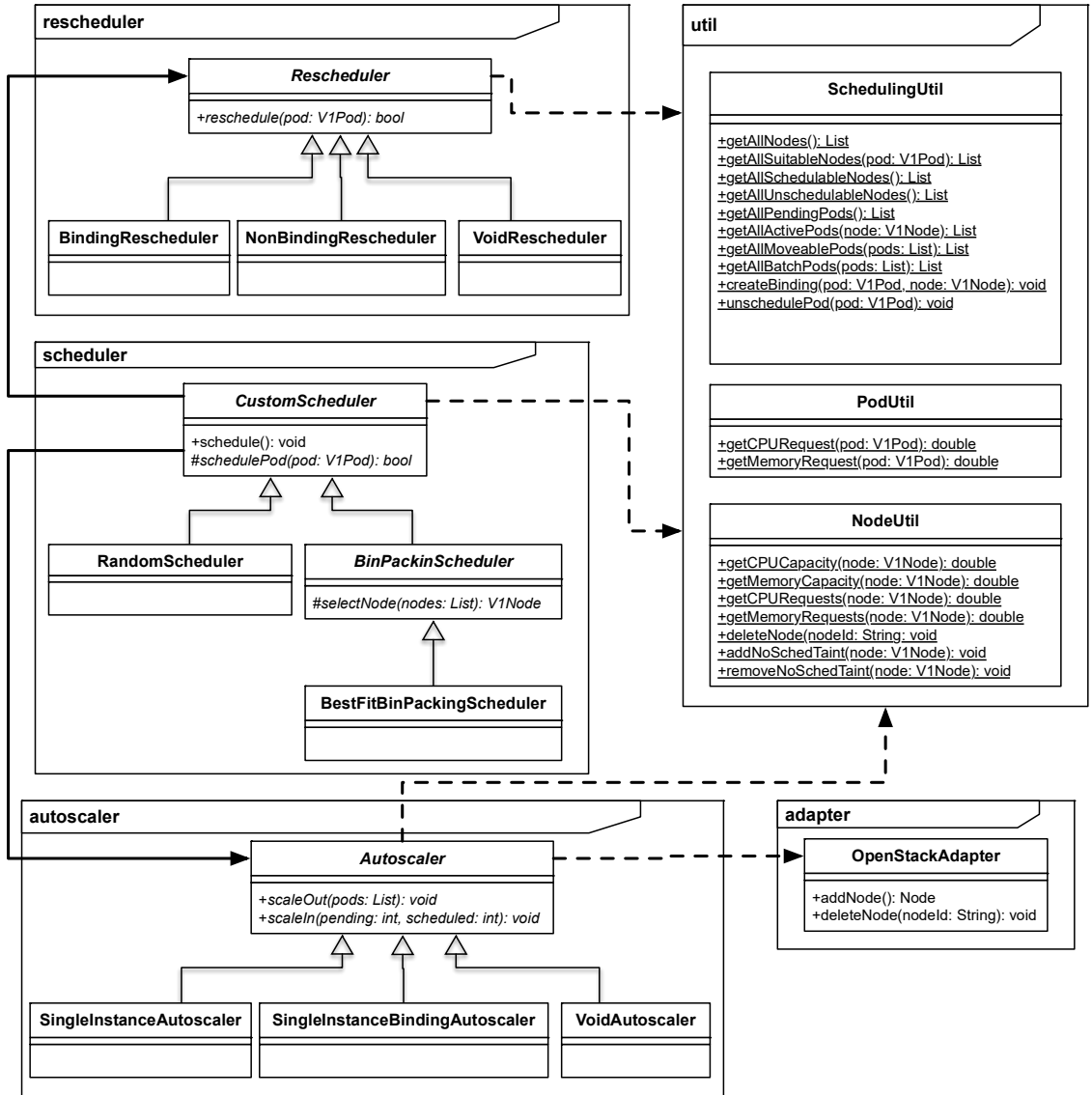

**Figure 2.** Class diagram for the prototype system

## 5. Design and Implementation

The prototype system has been implemented in Java. It runs as an independent process and hence does not need to be deployed in a node belonging to a Kubernetes cluster. As depicted in Figure 1, the system interacts with Kubernetes through the API. Figure 2 shows a high-level class diagram for the prototype system. We have included classes, methods, and attributes that are relevant to the discussion presented in this work. The classes in the *util* package are the ones interacting directly with Kubernetes while the rest of the classes contain the actual logic of the algorithms.

## 5.1. Kubernetes-specific Workload

Long running tasks are defined as Kubernetes deployments. A deployment creates a replica set that spawns the required number of replicated pods. A pod defines the actual application or task. Pods are recreated automatically when a replica fails. For the initial instance of this work, we assume deployments with a single replica. Furthermore, long running services defined as moveable must include a label indicating this (key: rescheduling, value: moveable). CPU and memory requests must be specified as well as CPU and memory limits, which should be equal to the requests to ensure the pods are assigned to the Kubernetes guaranteed QoS class. An example of the YAML file defining a deployment of a single replica of an *nginx* service is shown in Figure 3.

The same rules regarding requests/limits and replicas apply to batch jobs. However, these are defined using Kubernetes jobs as opposed to deployments, which allow the definition of tasks that run to completion. Furthermore, jobs cannot be labeled as moveable and instead must be labeled with key type and value batch. A sample YAML defining a job is depicted in Figure 4.

```yaml
apiVersion: apps/v1
kind: Deployment
metadata:
  generateName: nginx-
spec:
  replicas: 1
  selector:
    matchLabels:
      app: nginx
  template:
    metadata:
      name: nginx
      labels:
        app: nginx
        rescheduling: moveable
    spec:
      schedulerName: customScheduler
      containers:
      - name: nginx
        image: nginx
        ports:
        - name: http
          containerPort: 80
        resources:
          requests:
            memory: "1.4Gi"
            cpu: "100m"
          limits:
            memory: "1.4Gi"
            cpu: "100m"
```

**Figure 3.** YAML file defining a deployment of a single nginx service replica

```
apiVersion: batch/v1
kind: Job
metadata:
  generateName batch-
spec:
  template:
    metadata:
      labels:
        type: batch
    spec:
      schedulerName: customScheduler
      restartPolicy: Never
      containers:
        - name: batch-container
          image: busybox
          command: ['sh', '-c', ./batchJob']
          resources:
            requests:
              memory: "1Gi"
              cpu: "100m"
            limits:
              memory: "1Gi"
              cpu: "100m"
```

**Figure 4.** YAML file defining a batch job

## 6. Cost-Efficient Scheduling and Autoscaling Algorithms

This section describes the algorithms that were proposed and implemented in the prototype system. The design of the framework is flexible as it allows for any combination of scheduling, rescheduling, and scheduling to be used when launching the system.

### 6.1. Scheduling

*Best Fit Bin Packing Scheduler*

The general bin packing problem can be defined as follows. Given n items of different weights and bins each of capacity c, assign each item to a bin such that number of total used bins is minimized. There are multiple greedy solutions designed to solve the online version of this problem, that is, instances where items arrive one at a time and must be placed in a bin before considering the next item. One of these solutions is referred to as best fit, which places items in the fullest bin that can still accommodate an item without the bin capacity being exceeded.

For the scheduling problem addressed by this work, an item corresponds to a pod and a bin to a cluster node. Since pods are continuously arriving for execution, we consider the problem to be online as well. Hence, this scheduler places incoming pending pods in the node with the smallest amount of available resources that satisfy the pod requirements. This problem however is not one dimensional as pods require two different resources to be available, CPU and memory. We address this by first filtering nodes that have enough CPU available and then selecting from this list the node that has the least amount of available memory that is at least as large as the memory request of the incoming pod. The reason for this is two-fold. First, our goal is to keep the algorithm simple and capable of making fast decisions. Second, we consider CPU to be a compressible resource; that is, it's use can be throttled if the requested amount is exceeded or the node is overloaded. Memory on the other hand is non-compressible in that its use cannot be throttled. Stopping a pod from using excess memory can only be done by killing the pod and relieving memory pressure from an overloaded node can only be achieved by evicting pods currently deployed on the node.

```
Algorithm 2
Best Fit Bin Packing Scheduler
```
```
Input: Pending pod p
Output: true if successful, false otherwise

nodes = getAllSuitableNodes(p)
node = node with least amount of available RAM in nodes
if node is not null
  createBinding(node, p)
  return true
return false
```

## 6.2. Rescheduling

**Void Rescheduler**
The void rescheduler will simply ignore rescheduling request and hence is equivalent to a system without rescheduling capabilities.

**Non-binding Rescheduler**
This rescheduler aims to evict moveable pods from a node if and only if i) the moveable pods can be rescheduled somewhere else and ii) by evicting the moveable pods, the node now has enough resources to place an unschedulable pod. This is however only attempted if the unschedulable pod has been in a pending state for at least *max_pod_age*. This with the aim of reducing the number of unnecessary rescheduling and autoscaling decisions as it gives batch jobs the chance to complete and hence make room for the unschedulable pod.

Based on a best fit heuristic and on the same assumptions regarding compressible and non-compressible resources as the best fit bin packing scheduler, the non-binding rescheduler achieves its goals in the following way. First nodes are filtered to obtain only those that have enough available CPU to fit the unschedulable pod. These are then sorted ascendingly based on the amount of memory they have available. Then, for each of these nodes, if one or more moveable pods can be scheduled on a different node in the cluster and the amount of memory freed on the node is sufficient to execute the unschedulable pod, then the moveable pods are evicted and these and the unschedulable pod are left in the pending queue so that they can be placed by the scheduler in the next cycle (i.e., non-binding).

```
Algorithm 3
Non-binding Rescheduler
```
```
Input: Unschedulable pod p

If age of p is greater than or equal to max_pod_age
  nodes = getAllNodesWithEnoughCPU(p)
  sort nodes descending based on available memory
  for each node in nodes
    if the node has moveable pods running
      sort moveable pods descending based on requested memory
      for each moveable pod
        if the pod can be placed somewhere else
          add pod to list of pods to evict
          update freed memory
          if enough memory has been freed
            evict pods
            success = true
            break
      if success
        break
```

**Binding Rescheduler**

The binding rescheduler works in a similar way to the non-binding one. The only difference is that instead of leaving evicted and unschedulable pods in the pending queue, they are actually placed on their corresponding nodes by the rescheduler.

**Algorithm 4**
**Binding Rescheduler**

```
Input: Unschedulable pod p

If age of p is greater than or equal to max_pod_age
  nodes = getAllNodesWithEnoughCPU(p)
  sort nodes descending based on available memory
  for each node in nodes
    if the node has moveable pods running
      sort moveable pods descending based on requested memory
      for each moveable pod
        if the pod can be placed somewhere else
          add <pod, newNode> to map of pods to evict
          update freed memory
        if enough memory has been freed
          evict moveable pods from node
          for each pod in pods to evict
            createBinding(pod, newNode)
          success = true
          break
    if success
      break
```

### 6.3. Autoscaling

*Void Autoscaler*

The void autoscaler will simply ignore the scale out and scale in requests and hence is equivalent to a system without autoscaling capabilities.

*Simple Autoscaler*

The simple autoscaler will consider launching a new instance of a predefined type whenever the scale out operation is invoked. The number of instances launched is capped to one every *provisioning_interval*. The motivation behind this limit is based on the following observation. Unschedulable pods are likely to be found in batches. That is, if there are insufficient resources to deploy one pod, there may be insufficient resources to deploy the next pending pod in the queue. Hence, scaling out requests are likely to be made several times during the same scheduling cycle. This may lead to an excessive number of instances being launched which may end up being underutilized as a single one may have sufficed to execute the unschedulable pods. In fact, we set the *provisioning_interval* based on an estimate of the instance provisioning delay (i.e., the time it takes for the VM to boot and join the K8s cluster) plus a small contingency value. Notice however that this parameter is configurable by users.

**Algorithm 5**
**Simple Autoscaler – Scale Out**

```
Input: Unschedulable pod p

if elapsed time since last instance launched >= provisioning_interval
  launch new instance
  update last instance launch time
else
  ignore scale out operation
```

Regarding the scale in operation, it occurs only when the scheduling cycle was successful in placing all the pending pods in the queue. Also, only those nodes that were created dynamically (e.g., by scaling out) are considered to be shut down. If these conditions are met, firstly, any idle node with no pods running is deprovisioned. Secondly, any node that has only moveable pods that can be placed in any other node in the cluster is also shut down. Thirdly, any node that has a combination of moveable and batch pods is tainted as unschedulabe if all of its moveable pods can be placed on different nodes. If this is the case, moveable pods are evicted and left to be recreated by Kubernetes and placed by the scheduler in the next cycle. Because the node has been tainted as unschedulable, schedulers will avoid placing pods on the node unless strictly necessary (i.e., no other untainted node has sufficient capacity). In this way, the node has the potential of becoming idle once the batch jobs running in it complete and hence should be deprovisioned on a subsequent scaling in cycle.

**Algorithm 6**
**Simple Autoscaler – Scale In**

```
if all pending pods were successfully scheduled in the last cycle
   shutdown any nodes that are empty and were autoscaled
   nodes = getAllSchedulableNodes()
   for each node in nodes
      if all pods are moveable
         if all pods can be placed on other nodes
            delete node
            let Kubernetes recreate the pods
            let the scheduler place the recreated pods
      if some pods are moveable and some pods are batch jobs
            if all moveable pods can be placed on other nodes
               delete all moveable pods
               let Kubernetes recreate the pods
               let the scheduler place the recreated pods
               taint the node as unschedulable
```

*Single Instance Binding Autoscaler*
The heuristic is similar to that of the single instance autoscaler. However, this autoscaler aims to address a key disadvantage of the non-binding autoscaler. When unschedulable pods are not bound or associated in any way to autoscaled nodes, multiple requests to autoscale triggered by the same unschedulable pod may be invoked. This may lead to multiple instances being launched unnecessarily.

This autoscaler keeps track of pods and the scaled-out nodes that have been launched for them. In fact, the autoscaler keeps a list of pods that could potentially run on a node that is being provisioned. Requests to scale out made for a pod that is already associated with a node that is in the process of booting up are ignored. Once the node has been provisioned and has joined the cluster, the autoscaler is notified so that the pod is no longer associated to any node. The scheduler is responsible then for placing the pod on an available node in the next cycle, this node is likely to be the newly provisioned one, but this is not mandatory. If the node was filled with other pending nodes by the scheduler, then a request to scale out for the pod will be triggered and the process repeated.

To further prevent unnecessary nodes being launched, the binding autoscaler takes into consideration nodes that have been provisioned but have not joined the cluster yet when deciding whether to launch a new instance or not. When a request to scale out is made for an unschedulable pod, the autoscaler scans the list of nodes that are in the process of being provisioned. If there is still room in one of these nodes for the pending pod, then the pod is added to the list of pods that could potentially run on that node and no new instance is launched.

**Algorithm 7**
**Simple Binding Autoscaler – Scale Out**

```
Input: Unschedulable pod p

if pod has not been assigned to a provisioning node
   if pod can fit in one of the nodes being provisioned
      assign pod to provisioning node
   else
      launch a new node and assign the pod to the node
when new node is provisioned
   unassign any pod associated to the node
   remove node from list of provisioning nodes
   let the scheduler place pods in the new node
```

## 7. Performance Evaluation

### 7.1. Testbed and Workload

To demonstrate the potential capabilities of the proposed framework and the benefits of autoscaling, we performed a set of validating experiments. We modelled synthetic workloads composed of batch jobs and long running services as presented in Table 1. Three different workloads with varying inter-arrival job times and combination of jobs and services were used as illustrated in Table 2. To generate these workloads, jobs were selected at random with equal probability and the delay between them was sampled from an exponential distribution with different means. For the bursty workload, a mean of 10 seconds was used, the aim is to simulate jobs arriving at a high rate. For the slow workload, a mean of 60 seconds was used. The mixed workload combines both a bursty and a slow job arrival rate. It was generated by splitting the workload into periods, each being either bursty or slow. The first period was chosen at random and subsequent ones were alternated. The number of jobs in each period was chosen at random with the minimum number of jobs in a period being 10.

The jobs do not actually use the requested resources; however, this does not affect the obtained results as the purpose of the requests is to validate the functionality and performance of the scheduler.

**Table 1.** Types of jobs used in the evaluation

|  | Name | Task | Memory Requests | CPU Requests |
|---|---|---|---|---|
| **Batch Jobs** | batch_small | sleep 5 min | 0.3Gi | 100m |
|  | batch_med | sleep 10 min | 0.6Gi | 200m |
|  | batch_large | sleep 15 min | 0.9Gi | 300m |
| **Long-running Services** | service_small | nginx server | 1Gi | 100m |
|  | service_med | nginx server | 1.4Gi | 200m |
|  | service_large | nginx server | 2.359Gi | 300m |

**Table 2.** Workloads used in the evaluation

| Workload Name | Exponential Dist. Mean (sec) | batch | | | service | | | Total Batch | Total Service | Total |
|---|---|---|---|---|---|---|---|---|---|---|
| | | small | med | large | small | med | large | | | |
| Bursty | 60 | 10 | 8 | 5 | 6 | 12 | 9 | 23 | 27 | 50 |
| Slow | 10 | 17 | 11 | 4 | 6 | 7 | 5 | 32 | 18 | 50 |
| Mixed | 60 slow, 6 burtsy | 6 | 7 | 9 | 7 | 11 | 10 | 22 | 28 | 50 |

The Kubernetes cluster was deployed on Nectar [19], an Australian research cloud based on Openstack. The VM specifications used to deploy the Kubernetes master and worker nodes are depicted in Table 3. The custom architectural components were deployed outside Nectar on a MacBook Pro with a 2.9 GHz Intel Core i7 processor and 8 GB of RAM and the Kubernetes version used was 1.10. Since Nectar does not charge for the use of resources, we estimate the cost of each approach based on the billing model of existing cloud providers. In particular, we assume a per-second billing of $0.011 for each worker based on Microsoft Azure's general purpose *B2S* instance type, with any partial use being rounded up to the nearest second. Although Microsoft bills per minute, we assume a per-second billing due to the nature and scale of the experiments without impacting the significance of the results. Table 4 depicts the values used for the algorithm-specific parameters.

**Table 3.** Master and worker nodes VM specifications

|  | VM Type | # of vCPUs | RAM | Operating System |
|---|---|---|---|---|
| Master | m2.medium | 2 | 6 GB | Ubuntu 17.01 |
| Worker | m2.small | 1 | 4 GB | Ubuntu 17.01 |

**Table 4.** Algorithm-specific parameter values used in the evaluation

| Parameter | Value | Algorithm |
|---|---|---|
| VM price per second | $0.011 | N/A |
| max_pod_age | 1 min | Rescheduler |
| provisioning_interval | 1 min | Autoscaler |

## 7.2. Results

First, we evaluate the performance of each of the rescheduling and autoscaling heuristics in terms of cost and scheduling duration. We define scheduling duration as the time elapsed from the moment the first job is submitted and the moment the last batch job completes its execution. The cost is estimated based on the amount of time each VM was provisioned for; that is, from the moment a request for provisioning was placed to the cloud provider until the moment a deprovisioning request was placed. For static nodes, the cost was estimated based on the total scheduling time of the workload. Finally, only the rescheduling and autoscaling policies vary throughout the experiments, the scheduler (best fit bin packing) is kept constant.

Figure 3 presents the cost and total scheduling duration for each of the scheduler/autoscaler combinations and each of the workloads. For the mixed workload, the lowest cost and scheduling duration is obtained by the Non-binding Rescheduler and Binding Autoscaler (NBR-BAS). In fact, the binding autoscaler combined with any of the rescheduler always leads to the lowest cost and in 2 out of 3 cases, to the lowest scheduling duration. This emphasizes on the importance of having heuristics in place that aim to avoid unnecessary scaling out operations. While the non-binding autoscaler blindly provisions a new instance if a pod cannot be placed, the binding autoscaler considers instances that are currently being provisioned as potential hosts for the incoming pod before provisioning a new node. As for the difference in performance between the binding and non-binding reschedulers, it seems to be a better option to allow the scheduler to place all pending pods as opposed to trying to replicate the job of the scheduler in the rescheduler.

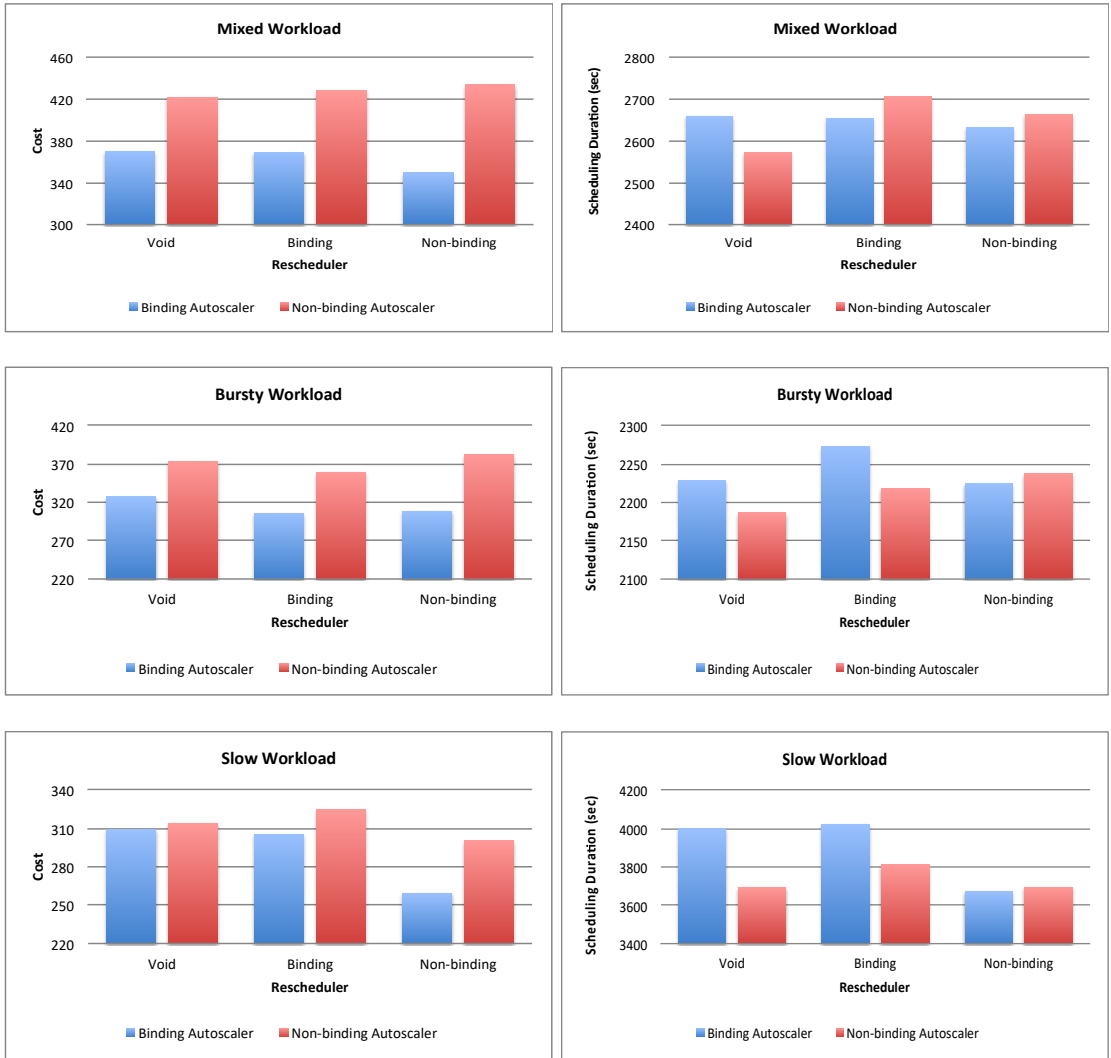

**Figure 3.** Cost and scheduling duration for the three different workloads and six different algorithms.

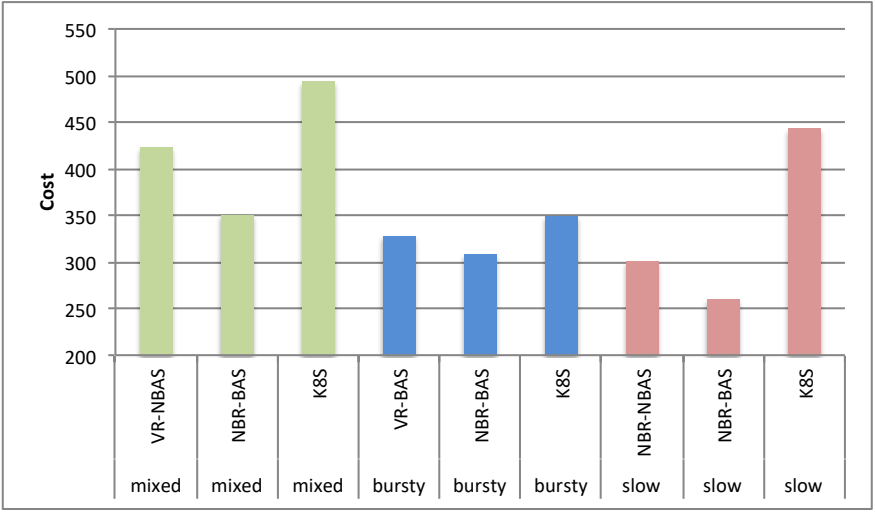

**A**. Cost of execution - default K8S vs others

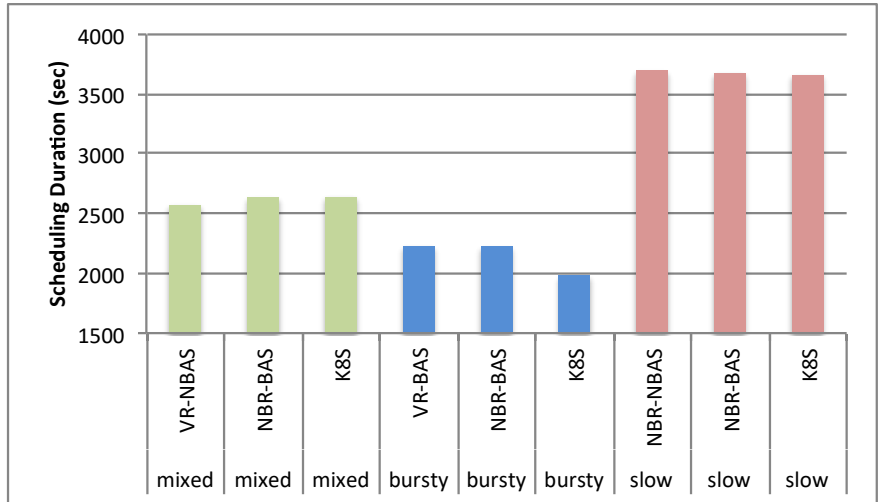

**B**. Scheduling duration - default K8S vs others

**Figure 4.** Cost and scheduling duration for the k8s scheduler and the best performing rescheduler/autoscaler combinations for different workloads (mixed, bursty, and slow).

The results in terms of cost for the bursty workload are similar to those of the mixed workload. Once again, the binding autoscaler leads to smaller number of resources used and hence lower costs. However, because resources are provisioned less frequently and in a more conservative manner and because the rate of job arrival is so high, jobs may have to wait longer before being placed and this leads to higher scheduling durations. For this particular workload, the existence of a rescheduler does not seem to make a significant impact in terms of cost and scheduling time. Perhaps rescheduling may have only delayed the provisioning of VMs and the scale of the experiments do not allow for the effects of these to be seen. However, in Table 5 we show how the actual utilization of resources does improve, although slightly for the reason above, when using a rescheduler.

For the slow workload, the best performing algorithm combination is once again NBR-BAS for both cost and scheduling duration. In this case, the difference in both metrics between this approach and the Void Rescheduler and Binding Autoscaler (VR-BAS) is significant and hence the benefits of rescheduling jobs to better utilize resources can be clearly appreciated.

To further understand the benefits of the proposed approaches, we also compare their performance to that of the default Kubernetes scheduler. We give each of the rescheduler/autoscaler combinations a score based on their cost and scheduling duration performance, where each of these has the same weight. We compare the Kubernetes scheduler (K8S) to the two algorithms with the lowest score (i.e., best performance) for each of the workloads. The results are depicted in Figure 4. The cost and scheduling time for K8S is estimated by running each workload on the minimum number of static nodes in which K8S can successfully place and execute all the jobs. In rescheduler/autoscaler algorithms combinations, VR corresponds to Void Rescheduler, NBR to Non-binding Rescheduler, NBAS to Non-binding Autoscaler, and BAS to Binding Autoscaler.

For every workload, the cost associated with the Kubernetes scheduler is considerably higher than the cost associated with the proposed approaches. The maximum reduction in cost is achieved for the slow workload, with NBR-BAS achieving a cost reduction of over 58% when compared to the Kubernetes default scheduler. Regarding the scheduling duration, the performance of our algorithms is only slightly worse than that of K8S. Considering the benefits obtained in terms of cost, we consider this to be a very reasonable trade-off.

**Table 5.** Scheduling performance and resource utilization for each of the rescheduler/autoscaler combinations.

| Workload | Rescheduler | Autoscaler | Median Scheduling Time (sec) | Average RAM Req/Cap | Average CPU Req/Cap | Average # pods/node |
|---|---|---|---|---|---|---|
| **Mixed** | Void | Non-binding | **25.00** | 0.72 | 0.68 | 2.03 |
| | Binding | Non-binding | 37.00 | 0.73 | 0.67 | 1.99 |
| | Non-binding | Non-binding | 111.00 | 0.79 | **0.72** | 2.22 |
| | Void | Binding | 49.50 | **0.81** | **0.72** | 2.20 |
| | Binding | Binding | 26.00 | 0.77 | 0.70 | 2.14 |
| | Non-binding | Binding | 112.50 | 0.80 | **0.72** | **2.23** |
| **Slow** | Void | Non-binding | 14.00 | 0.79 | 0.72 | 2.41 |
| | Binding | Non-binding | 11.50 | 0.78 | 0.70 | 2.33 |
| | Non-binding | Non-binding | **10.00** | 0.83 | 0.74 | 2.52 |
| | Void | Binding | 12.50 | 0.82 | 0.72 | 2.44 |
| | Binding | Binding | 17.50 | 0.84 | 0.74 | 2.49 |
| | Non-binding | Binding | 12.50 | **0.86** | **0.76** | **2.62** |
| **Bursty** | Void | Non-binding | 422.50 | 0.65 | 0.69 | 2.11 |
| | Binding | Non-binding | **297.00** | 0.67 | 0.71 | 2.24 |
| | Non-binding | Non-binding | 467.00 | 0.62 | 0.67 | 2.04 |
| | Void | Binding | 418.00 | 0.70 | 0.72 | 2.28 |
| | Binding | Binding | 354.00 | 0.71 | 0.73 | 2.31 |
| | Non-binding | Binding | 388.50 | **0.73** | **0.74** | **2.36** |

Finally, in Table 5 we present a more detailed insight into the performance of each of the approaches. In particular, we present the median scheduling time, which corresponds to the median time spent by pods in a pending state. We also present the average RAM requests to RAM capacity ratio of each node. The average RAM requests of a node corresponds to the average of measurements taken every 20 seconds throughout the scheduling duration. These measurements are the sum of the memory requests of all the pods allocated to the node. The RAM capacity of a node is the maximum amount of memory that can be requested by all of the pods in the node. The average CPU requests to capacity ratio is also depicted in Table 5 and is estimated in the same way as the RAM ratio. The average number of pods deployed on a node is also shown.

The median scheduling time is the fastest for the slow workload and the NBR-NBAS. This is in line with what is expected of the algorithms as the non-binding autoscaler causes more resources to be provisioned and hence readily available for use as jobs arrive. Furthermore, a larger interarrival time between jobs means batch jobs can complete and hence space can be freed by the time a new job arrives. This is further supported by the large difference between the median times for the slow and bursty workloads. Jobs in the bursty workload have to wait for a considerably longer amount of time in pending state as resources cannot be provisioned fast enough. This due mainly to VM provisioning delays.

Regarding the RAM utilization ratio, all of the approaches using the non-binding autoscaler achieve the worst performance. This is because resources are overprovisioned and hence underutilized. Similar results were obtained for the CPU ratio. The benefits in terms of better utilization are clear for the NBR-BAS approach on the bursty workload, with it obtaining the highest ratios and average number of pods per node.

## 8. Discussion and Future Directions

There are various areas that could be further explored in order to better optimize the decisions made by the schedulers, reschedulers, and autoscalers. For instance, considering heterogeneous VMs could lead to a more efficient use of resources and decreased cost. This would enable for instance to dynamically provision resources with different pricing models to the virtual cluster in order to satisfy growing needs of the applications with minimum cost. For example, a customer-facing application should be placed on reserved instances that are leased for lower costs and longer periods of time while offering high availability. Batch jobs on the other hand could be placed on unreliable rebated resources, whose sudden termination will not disrupt the end user experience. The use of on-demand instances can be explored for applications with requirements in between where the availability is needed but they are not long-running services.

Implementing algorithms that are data-aware is another area worth exploring. This data-awareness may refer to either the container images or application data. Considering ephemeral storage as a resource in addition to memory and CPU can also further improve the proposed system and solutions. Extending the definition of moveable pods to a more robust one could also make the framework more appealing to users with more stringent requirements. For instance, containers may be defined as moveable if there are more than one replicas of a given pod currently running and the application can tolerate one of the replicas being shut down and restarted (e.g., stateless services or applications using persistent volumes).

Application QoS management is another improvement worthwhile exploring. Applications have specific QoS requirement, for instance, long-running services commonly have to serve a minimum number of requests per time unit or have stringent latency requirements. Batch jobs on the other hand can have a deadline as a time constraint for their execution or may need to be completed as fast as possible. For the first scenario, many systems offer a basic autoscaling mechanism. It monitors the CPU utilization of a service, and if a predefined threshold is exceeded, another instance of the service is launched. This however, is a baseline approach to autoscaling and integrating more sophisticated approaches to container-based management systems is required. For batch jobs, orchestrating them and assigning them to resources so that their QoS are met is another open research area. This feature is not present in any open source system and support for heterogeneous QoS constraints is still unexplored.

Resource consumption estimation can be used to predict and estimate the amount of resources a container consumes at different points in time, as opposed to relying simply on the amount of resources requested for a particular container. The reason is twofold. Firstly, resource requests are usually misestimated and overestimated by users. Secondly, the resource consumption of a task is likely to vary over time, with the peak consumption spanning only over a fraction of its lifetime. Both scenarios lead to resources that are reserved but are idle most of the time and hence lead to the cluster being underutilized. By monitoring and estimating the resource consumption of containers, better oversubscription and opportunistic scheduling decisions can be made by the system. In the proposed system prototype, this can be achieved by making use of Kubernetes metrics server, an InfluxDB to store the timeseries of the collected consumption data, and an online prediction algorithm designed for streaming data such as Hierarchical Temporal Memory [18].

## 9. Conclusions

In this work, we have proposed the integrated use of schedulers, autoscalers, and reschedulers as a mechanism to make container orchestration systems cloud-aware. In this, the scheduler optimizes the initial placement of containers, the autoscaler enables the current demand for resources to be met

and underutilized or idle nodes to be shutdown improve the system's utilization and hence reduce cost, and finally the rescheduler allows for the initial placement of containers to be revised at runtime to reduce fragmentation and consolidate loads to encourage better resource utilization. A prototype system was developed as an extension to Kubernetes, a widely used open source container management system. Various rescheduling and autoscaling mechanism were proposed, implemented, and evaluated. Overall, we found rescheduling and autoscaling have clear benefits in terms of minimizing the cost of computation and enhancing the resource efficiency. We also identified and discussed future research directions.


**Acknowledgments**

This work supported through a collaborative research agreement between the University of Melbourne and Samsung Electronics (South Korea) as part of the Samsung GRO (Global Research Outreach) program. We thank Minxian Xu for his help with improving the quality of paper.



**References**

[1] Docker. https://www.docker.com. Accessed on June2018.
[2] D. Bernstein, Containers and Cloud: From LXC to Docker to Kubernetes, IEEE Cloud Computing, vol. 1, no. 3, pp. 81-84, 2014.
[3] Verma A, Pedrosa L, Korupolu M, Oppenheimer D, Tune E, Wilkes J. Large scale cluster management at Google with Borg. In Proceedings of the 10th European Conference on Computer Systems (EuroSys 2015), Bordeaux, France — April 21 - 24, 2015. p. 18.
[4] Hightower K, Burns B, Beda J, Kubernetes: Up and Running: Dive Into the Future of Infrastructure, O'Reilly Media, Inc., 2017.
[5] V. Medel, O. Rana, J. Á. Bañares and U. Arronategui, Modelling Performance & Resource Management in Kubernetes, In Proceedings of the *2016 IEEE/ACM 9th International Conference on Utility and Cloud Computing (UCC)*, Shanghai, 2016, pp. 257-262.
[6] N. Naik, Building a virtual system of systems using docker swarm in multiple clouds, In Proceedings of the 2016 IEEE International Symposium on Systems Engineering (ISSE), Edinburgh, 2016, pp. 1-3.
[7] Hindman B, Konwinski A, Zaharia M, Ghodsi A, Joseph AD, Katz RH, et al. Mesos: A Platform for Fine-Grained Resource Sharing in the Data Center. In Proceedings of the USENIX Symposium on Networked Systems Design and Implementation (NSDI); 2011. p. 22–22.
[8] Marathon. https://mesosphere.github.io/marathon. Accessed on June2018.
[9] R. DelValle, G. Rattihalli, A. Beltre, M. Govindaraju and M. J. Lewis, Exploring the Design Space for Optimizations with Apache Aurora and Mesos, In Proceedings of the 2016 IEEE 9th International Conference on Cloud Computing (CLOUD), San Francisco, CA, 2016, pp. 537-544.
[10] Vavilapalli VK, Murthy AC, Douglas C, Agarwal S, Konar M, Evans R, et al. Apache Hadoop YARN: Yet Another Resource Negotiator. In Proceedings of the 4th ACM Annual Symposium on Cloud Computing; 2013. p. 5.
[11] Containerd. https://containerd.io. Accessed on August 2018.
[12] Frakti. https://github.com/kubernetes/frakti. Accessed on August 2018.
[13] CRI-O. http://cri-o.io. Accessed on August 2018.
[14] runc. https://github.com/opencontainers/runc. Accessed on August 2018.
[15] etcd. https://coreos.com/etcd. Accessed on June 2018.
[16] Schwarzkopf M, Konwinski A, Abd-El-Malek M, Wilkes J. Omega: flexible, scalable schedulers for large compute clusters. In Proceedings of the 8th ACM European Conference on Computer Systems; 2013. p. 351–364.
[17] Kubernetes, Building Large Clusters. https://kubernetes.io/docs/admin/cluster-large. Accessed on June 2018.
[18] Y. Cui, S. Ahmad, J. Hawkins, Continuous online sequence learning with an unsupervised neural network model, arXiv:1512.05463, 2015.
[19] Nectar. https://nectar.org.au/. Accessed on August 2018.
[20] Jha D, Garg D, Jayaraman P, Buyya R, Li Z, and Ranjan R, A Holistic Evaluation of Docker Containers for Interfering Microservices, Proceedings of the 2018 IEEE International Conference on Services Computing (SCC 2018, IEEE CS Press, USA), San Francisco, USA, July 2-7, 2018.



[21] Sossa M and Buyya R, Container-based Cluster Orchestration Systems: A Taxonomy and Future Directions, Software: Practice and Experience (SPE), 2018, DOI: https://doi.org/10.1002/spe.2660
[22] Xu X, Yu H, and Pei X, A novel resource scheduling approach in container based clouds, in Proceedings of 2014 IEEE 17th International Conference on Computational Science and Engineering (CSE), Dec 19 2014, p. 257-264.
[23] Zhang H, Ma H, Fu, G, Yang, X, Jiang, Z, and Gao Y, Container based video surveillance cloud service with fine-grained resource provisioning, in Proceedings of 2016 IEEE 9th International Conference on Cloud Computing (CLOUD), pp. 758-765.
[24] Kaewkasi C, and Chuenmuneewong K, Improvement of container scheduling for docker using ant colony optimization, in Proceedings of 2017 9th IEEE International Conference on Knowledge and Smart Technology (KST), 2017, pp. 254-259.
[25] Yin L, Luo J and Luo H, Tasks Scheduling and Resource Allocation in Fog Computing Based on Containers for Smart Manufacturing, IEEE Transactions on Industrial Informatics, 2018, 14(10): pp. 4712-4721.
[26] Guerrero C, Lera I, and Juiz C, Genetic algorithm for multi-objective optimization of container allocation in cloud architecture. Journal of Grid Computing, 2018, 16(1), pp. 113-135.
[27] Kehrer S, and Blochinger W, TOSCA-based container orchestration on Mesos. Computer Science-Research and Development, 2018, 33(3-4), 305-316.
[28] Xu M, Toosi A, and Buyya R, iBrownout: An Integrated Approach for Managing Energy and Brownout in Container-based Clouds, IEEE Transactions on Sustainable Computing (T-SUSC), DOI:10.1109/TSUSC.2018.2808493, 2018, pp. 1-14.